\documentclass[conference]{IEEEtran}
\IEEEoverridecommandlockouts
\usepackage{cite}
\usepackage{amsmath,amssymb,amsfonts}
\usepackage{algorithmic}
\usepackage{graphicx}
\usepackage{textcomp}
\usepackage{xcolor}
\usepackage{comment}
\usepackage{hyperref}

\newcommand{\publicationNotice}{
    \begin{center}
        \fbox{
            \parbox{0.9\textwidth}{
                \textbf{Published Article:} This article has been published by IEEE in the Proceedings of 2023 49th Euromicro Conference on Software Engineering and Advanced Applications (SEAA), Durres, Albania, 2023, pp. 5-12. \textbf{DOI:} \href{https://doi.org/10.1109/SEAA60479.2023.00011}{10.1109/SEAA60479.2023.00011}\\
                © 2023 IEEE.  Personal use of this material is permitted.  Permission from IEEE must be obtained for all other uses, in any current or future media, including reprinting/republishing this material for advertising or promotional purposes, creating new collective works, for resale or redistribution to servers or lists, or reuse of any copyrighted component of this work in other works.
            }
        }
    \end{center}
}

\newboolean{showcomments}
\setboolean{showcomments}{true} 
\ifthenelse{\boolean{showcomments}}
  {
		\newcommand{\nbb}[2]{
		\fcolorbox{black}{yellow}{\bfseries\sffamily\scriptsize#1}
		{\sf$\blacktriangleright$\textcolor{blue}{\textit{#2}}$\blacktriangleleft$}
		}
		
		\newcommand{\remarks}[1]{\color{red}[#1]\color{black}}

		\newcommand{\del}[1]{\textcolor{red}{\sout{#1}}} 
  }
  {
		\newcommand{\nbb}[2]{}
		\newcommand{\remarks}[1]{}

		\newcommand{\del}[1]{} 
  }

\newcommand{\ali}[1]{\nbb{Ali}{#1}}

\def\BibTeX{{\rm B\kern-.05em{\sc i\kern-.025em b}\kern-.08em
    T\kern-.1667em\lower.7ex\hbox{E}\kern-.125emX}}
\begin{document}
\publicationNotice


\title{On STPA for Distributed Development of Safe Autonomous Driving: An Interview Study
\thanks{Funded by Sweden's Innovation Agency, Diarienummer: 2021-02585}}

\author{\IEEEauthorblockN{Ali Nouri}
\IEEEauthorblockA{\textit{Volvo Cars} \\
Gothenburg, Sweden \\
ali.nouri@volvocars.com}
\and
\IEEEauthorblockN{Christian Berger}
\IEEEauthorblockA{\textit{University of Gothenburg, Sweden} \\
\textit{Department of Computer Science and Engineering}\\
christian.berger@gu.se}
\and
\IEEEauthorblockN{Fredrik Törner}
\IEEEauthorblockA{\textit{Volvo Cars} \\
Gothenburg, Sweden \\
fredrik.torner@volvocars.com}
}

\maketitle

\begin{abstract}
Safety analysis is used to identify hazards and build knowledge during the design phase of safety-relevant functions. This is especially true for complex AI-enabled and software intensive systems such as Autonomous Drive (AD). 
System-Theoretic Process Analysis (STPA) is a novel method applied in safety-related fields like defense and aerospace, which is also becoming popular in the automotive industry. However, STPA assumes prerequisites that are not fully valid in the automotive system engineering with distributed system development and multi-abstraction design levels. 
This would inhibit software developers from using STPA to analyze their software as part of a bigger system, resulting in a lack of traceability. This can be seen as a maintainability challenge in continuous development and deployment (DevOps).
In this paper, STPA's different guidelines for the automotive industry, e.g. J31887/ISO21448/STPA handbook, are firstly compared to assess their applicability to the distributed development of complex AI-enabled systems like AD. Further, an approach to overcome the challenges of using STPA in a multi-level design context is proposed. By conducting an interview study with automotive industry experts for the development of AD, the challenges are validated and the effectiveness of the proposed approach is evaluated.
\end{abstract}


\section{Introduction}
Autonomous Drive (AD) is foreseen as a possible solution towards vision zero \cite{SEAACDDM}. However, arguing for the safety of such a system is hard due to the complexity of such a software-intensive system, and because of the use of novel technologies like machine learning. Verification and validation by testing \cite{sotif} is one of the methods to check if the function is safe and meets the expectations of stakeholders like Original Equipment Manufacturer (OEM), customer, and regulators. However, finding and fixing design mistakes after the development phase is not only costly but also delays the release of the product \cite{SAESTPA}. Moreover, there is a need for evidence to argue for the safety of a function \cite{sotif, ISO26262} since testing is not capable of identifying all design mistakes in safety-related systems.

Analysis methods such as FMEA, FTA, and STPA can be employed to spot design mistakes \cite{ISO26262} and performance limitations \cite{sotif} that potentially might lead to hazards during the early development phase.
STPA provides a systematic approach to identify and analyze unsafe behavior in complex systems like AD and is referenced for AD development by regulators like ``United Nations Economic Commission for Europe'' (UNECE) \cite{ALKS}.


\subsection{Background}
\begin{figure}
\centering
  \includegraphics[width= 0.5\textwidth]{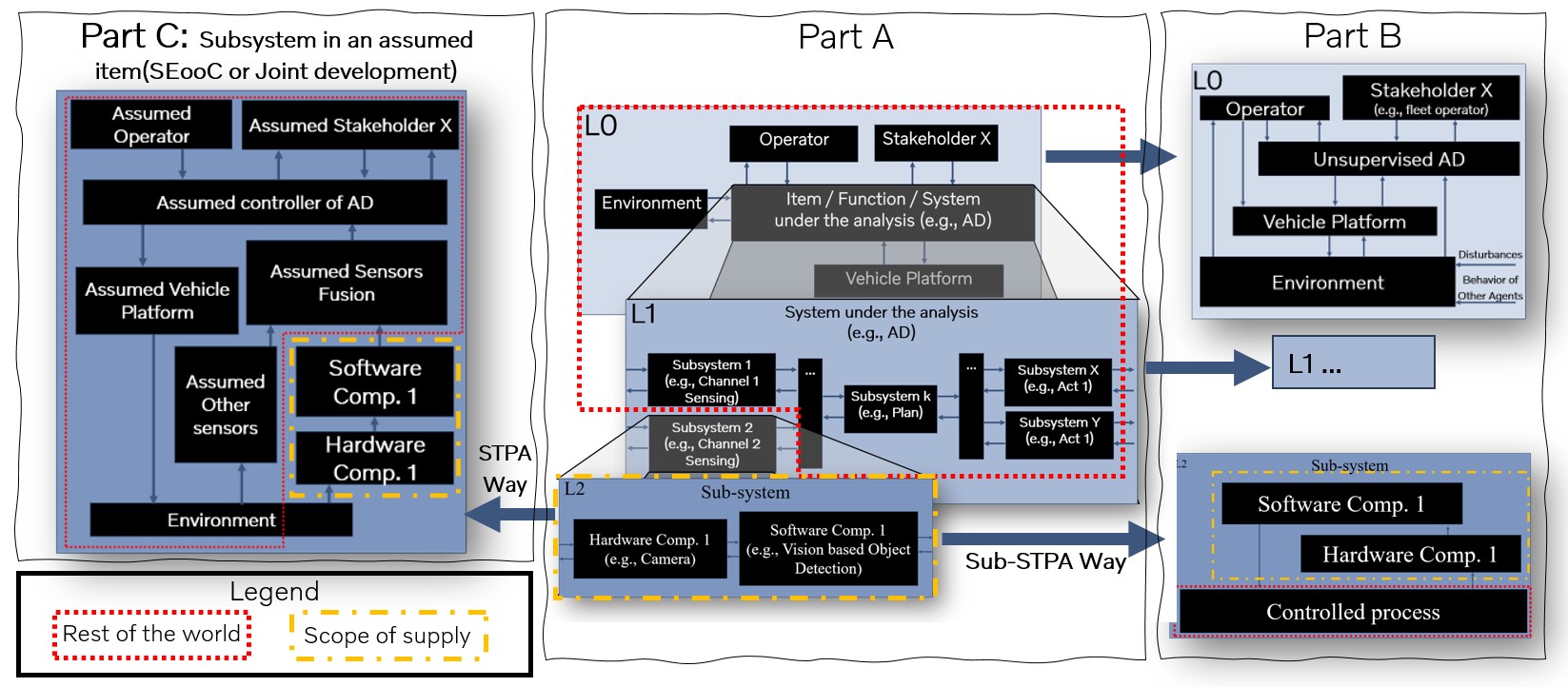}
  \caption{Part A illustrates the multi-abstraction level architectural design, beginning from the stakeholders' level (level 0) and progressing to the most intricate building blocks of the system (level 2). Part C illustrates the fundamental control structure of STPA for analyzing the L2 subsystem, which should be carried out by assuming the blocks outside the scope of supply (inside red dotted lines). Part B introduces an alternative approach proposed in this study.}
  \label{fig:abstraction}
\end{figure}

The architecture and requirements in the automotive domain follow well-defined modular and multi-abstraction architectural levels, which are an enabler for distributed development. Each modular element in each level is abstracted by merging or removing unnecessary characteristics \cite{SAESTPA} and assigned to internal or external teams for the design, implementation, and verification. Hence, it is important to clearly define each element in each abstraction level like the one presented in Fig.~\ref{fig:abstraction} Part A. 
Following the same design abstraction levels in all development activities including STPA analysis enables traceability, which leads to maintainability and understanding of the impact of changes in any element on the system.
By using a distributed development approach, the OEM can tap into the expertise of multiple suppliers and ensure that each aspect of the system is developed to the highest performance and quality especially for software-intensive and complex designs like AD. However, there is a need for confidentiality, necessitated by a combination of competitiveness, legality, and business factors. This leads to a situation where the development can become opaque, obscuring the details of the system's development between the parties involved in the development.

\subsection{Problem Domain \& Motivation}
\label{sec:ProblemDomain}
Despite STPA not providing specific guidance for managing the distributed development of automotive systems, the industry standards (e.g., ISO 21448) and regulations (e.g., UNECE R157) recommend its use. Therefore, it is important to consider adaptations of STPA to effectively address the challenges and leverage its advantages in the development process.

\subsection{Research Goal \& Research Questions}
The goal of research is to gather challenges and adapt the STPA method for the analysis of complex and safety-critical systems with multiple abstraction levels in a distributed development environment, with the following research questions:
\begin{description}
\item[RQ1:] What are the challenges of applying STPA in an automotive context with multi-abstraction levels in distributed development for OEM and suppliers?

\item[RQ2:] What do existing guidelines and publications propose for handling the multi-abstraction level designs in a distributed development environment for each involved party?

\item[RQ3:] How could STPA be modified to overcome the identified and confirmed shortcomings in RQ2 for each involved party?
\end{description}

\subsection{Contributions}
This work is conducted in the context of safety-related automotive system development in a distributed development environment. We identified the challenges for using STPA in complex systems like AD, which needs a multi-abstraction levels architecture in a distributed development environment, and validated them (RQ1). We study, compare, and analyze the main guidelines of the STPA in order to extract the existing instructions (RQ2). We proposed and validated an adaptation of STPA to address and overcome preconditions and specifics of the automotive system engineering process (RQ3). Both STPA for OEM and the proposed sub-system STPA (Sub-STPA) are mapped to ISO 26262 and ISO 21448 activities in each abstraction level and their traceability is demonstrated in Fig. \ref{fig:Traceability} to aim for practicality of the proposed method (RQ3).


\subsection{Scope}
This study is focusing on the context of Functional Safety, SOTIF, and safety aspects of cyber-security for the rapid development of safety critical AD functions. While selected challenges and ideas may be applicable to other contexts as well, we are only focusing on the automotive domain. 

This study's focus is not on comparing or integrating STPA with other methods. Rather, it aims to identify challenges and solutions for using STPA alone in validating complex systems like AD during distributed development. This enables subsystem developers to use STPA, achieving traceability of STPA elements to the most granular level.

\subsection{Structure of the Article}
In this article, the structure of the remaining sections is as follows:
In Sec.~\ref{sec:relatedwork}, we discuss related work followed by a description of our methodology
in Sec.~\ref{sec:methodology}. We identified the challenges and discuss proposed adaptations to the STPA in Sec.~\ref{sec:results_discussion} and discussed threats to validity in Sec.~\ref{sec:Threats}.
We conclude our findings in Sec.~\ref{sec:conlusion}.




\section{Related Work}
\label{sec:relatedwork}

Koelln et al. \cite{COMPAD2} applied STPA on an abstract AD architecture and analyzed the interaction of stakeholders like driver, legislature, and producer during Minimum Risk Maneuver. 
As part of the ``KI Absicherung - Safe AI for Automated Driving'' project, Celik et al. in \cite{asimsafecomp} investigate the use of the STPA approach for safety requirements elicitation in a Machine Learning based pedestrian collision avoidance system.
Abdulkhaleq et al. \cite{ASIMabd}  proposed an approach to integrate STPA into the software engineering process and raised the importance of using new methods like STPA over traditional methods like Fault Tree Analysis (FTA). Although Abdulkhaleq et al. in \cite{asimconti} acknowledge the importance of abstraction levels in AD system development, their setup involves a highly complex, granular level that cannot be decomposed into smaller modules and analyzed by a team of experts in a single area using STPA. Furthermore, this configuration is not adaptable for distributed development due to the necessity of confidentiality and protection of intellectual property.
Tran et al. in \cite{STPAFMEASW} explore the combined use of three hazard analysis methods including STPA to enhance software safety. However, they note that while STPA is effective in identifying hazards at system level, it may not be sufficient for analyzing more detailed levels such as subsystem or software. Therefore, the study highlights the need for other types of analysis such as FMEA to cover such additional abstraction levels. More over, Mallya et al. reported a lack of principles for STPA despite available guidelines, highlighting the need for attention to select the suitable abstraction level\cite{UsingSTPAinISO26262}. This indicates that the STPA method may not be suitable for all abstraction levels.
Although the ``STPA Handbook'' \cite{STPAhandbook} serves as the main reference for conducting STPA, it requires tailoring for its use in the automotive industry as it lacks guidance for distributed development and multi-abstraction levels. Similarly, while the guidelines from ``NHTSA'' \cite{ NHTSASOTIF}, ``ISO 21448'' \cite{sotif}, and SAE J3187 \cite{SAESTPA} provide valuable automotive-specific instructions, they also lack guidance on distributed development and multi-abstraction levels.


\section{Methodology}
\label{sec:methodology}

\textbf{For RQ1 and RQ3:} 
We conducted a semi-structured interview study to validate the identified challenges and proposed an approach (Sub-STPA) concerning the use of STPA in automotive system engineering. 
The questions were asked in a consistent order, with a yes/no or multiple-choice format for the first part, and an open-ended format for the second part that were designed to encourage the participants to reflect on their experiences and provide in-depth answers. This approach allowed participants to elaborate on their choices or provide alternative answers if needed, providing the researcher with new insights beyond the limitations of structured interviews.
The interview study was conducted with 14 participants from industry. The participants were selected based on their expertise in safety and hands-on experience with complex systems such as AD. The participants had on average 15 years of automotive experience and 12 years of safety experience, making them well-suited to provide valuable insights on the topic being studied. The study design was carefully planned and executed to ensure that the data collected was reliable and valid\footnote{Appendix providing all interview study details, \url{https://doi.org/10.5281/zenodo.8142674}}.


\underline{\emph{Interview Method:}} The interviews were conducted using Microsoft Teams from January to March 2023 and lasted between 60 and 120 minutes. The interviews were automatically transcribed and validated against the recorded interviews to ensure accuracy.
Explanation videos of terms, methods, and questions were pre-recorded and played during the interview to ensure that all participants received the same information and explanation. Although the interviewer was allowed to clarify if participants requested additional information that did not affect their answers. In case of questions that may influence the participants' answers, the interviewer refrained from answering and postponed the response until the end of the interview, when all answers had been finalized and the interview is finished.

\underline{\emph{Detailed Interview Protocol:}} A detailed interview protocol was designed to guide the interviewer and ensure consistency in the data collected. The protocol was finalized after the pilot study and before the first interview and included the goal of the interview, the steps to be taken before, during, and after the interview, and a list of questions to be asked in each step.

\underline{\emph{Questions:}} The questions asked during the interviews were designed to allow for open-ended discussion and encourage the participants to reflect on their experiences. The interview questions can be found in the provided appendix. 
In all multiple-choice questions, there was also the option of selecting ``none'', ``both'', or another option, which the participant was aware of and was not included in the choices. Questions 1-3 serve to validate the assumptions made by the researchers in this study regarding the challenges of using STPA in distributed development between OEMs and suppliers, particularly for subsystem suppliers. After introducing STPA and Sub-STPA, participants were asked question 4 regarding their preferred option (i.e., STPA or Sub-STPA) for analyzing a subsystem by a subsystem supplier.

\underline{\emph{Sample of the participants:}} The participants in this study were selected from a pool of safety experts in the automotive industry. To be eligible, the participants had to have at least five years of experience in safety-related topics and a background working in OEMs or suppliers. The participants were sourced through internal or external communities or LinkedIn. The study benefits from the fact that due to the distributed development environment in the automotive industry, the process of performing safety activities (e.g., safety analysis) is highly standardized, with joint reviews and collaborations with other OEMs and suppliers, and all parties following established standards. 

\underline{\emph{Pilot interview:}} A pilot interview was conducted prior to the main interviews to assess the effectiveness of the interview protocol. The results of the pilot study were not included in the final data analysis. This pilot study served to confirm the structure, relevance, effectiveness, and understandability of the questions asked during the interviews.

\underline{\emph{Data analysis:}} The transcripts of the interviews were validated against the recordings and then sent to the participants for validation. We performed the clustering of the answers if and only if the participant clearly agreed with one option, and then the interviewer confirmed the answer with the participant. 

\underline{\emph{Ethical Consideration:}} We followed the recommendations for ethical principles for software engineering interview studies as prescribed by Strandberg \cite{swethic} including anonymization, consent, and confidentiality. Each step of the interview was checked against the ethical checklist.

\textbf{For RQ2:} A literature survey of the existing technical literature and academic publications on the use of STPA in the automotive industry was conducted with a focus on multi-abstraction levels and distributed development environment within the context of functional safety and SOTIF. 
The dominating guidelines are selected 
that are relevant for industrial practitioners when developing safety critical functionality. We referenced the STPA handbook\cite{STPAhandbook}, the primary guideline for the method, along with automotive standards/guidelines from ISO, SAE, and NHTSA \cite{sotif, SAESTPA, NHTSASOTIF}.
We added results from screening relevant secondary literature that is discussing about and commenting on these
guidelines by including results from IEEE and ACM using the search term:

\begin{verbatim}
(((Autonomous OR Automated Or Driverless)
AND (Vehicle OR Drive OR Driving OR Car))
OR Self-driving) AND ((system theoretic
process analysis) OR STPA)
\end{verbatim}

These results were extended by extracting references from the industrial guidelines such as ISO 21448 as well
as by adopting snowballing for our results to expand on relevant secondary literature. Afterwards, the results
were selected based on their relevance to AD and STPA.
We used all papers to complement our own industrial
experience from working with these guidelines internally at an automotive OEM and from discussions conducted at ``2022 MIT STAMP Workshop''\footnote{Cf.~2022 MIT STAMP Workshop, \url{http://psas.scripts.mit.edu/home/2022-stamp-workshop-program/}}, where a public workshop was conducted to address ``STPA for Autonomous Vehicles Functions''.


\textbf{For RQ3:} We proposed an adaptation (Sub-STPA) that shall address the shortcomings of the STPA for distributed system engineering. We used an hypothetical example to illustrate the methodology. The ``Unsupervised AD'' function as an illustrating hypothetical example is used to present the method. The function must consider all relevant precautionary safety behaviors to maintain the vehicle in a safe and collision-free space. This is achieved using diverse sensor technologies including a vision-based object detection subsystem, which is developed by a tier two supplier.

\section{Results \& Discussion}
\label{sec:results_discussion}

As a result of the literature study in combination with the interview study, we identified several gaps and challenges in the existing STPA guidelines to be applicable for automotive safety-related function development in the aforementioned context.
We present automotive system engineering challenges faced by the OEM along with proposed STPA instructions to overcome those challenges in Sec. \ref{subsec:solution OEM}. In Sec. \ref{subsec:challenge supplier}, additional challenges faced by suppliers are presented, followed by a proposed adaptation of STPA for subsystems (Sub-STPA).

\subsection{OEM-level STPA Limitations and Proposed Adaptations:}

Firstly, Sec. \ref{subsubsec:challenge OEM}.1 introduces the challenges identified in this study. Then in Sec. \ref{subsubsec:challenge OEM}.2 the modified STPA process is presented to address the identified challenges.

\subsubsection{4.1.1 OEM-level Identified Challenges} Here, the limitations of STPA in automotive system engineering at OEM-level are presented to answer RQ1.
\label{subsubsec:challenge OEM}

\vspace{0.3cm}
\begin{tabular}{|p{0.9\columnwidth}|}
  \hline
  \textbf{Challenge 1- Mapping STPA steps to automotive process (supported by literature):} \\ 
  \hline
\end{tabular}
\vspace{0.1cm}

The differences in terminologies in the STPA steps with automotive standards such as ISO 26262 or 21448 may cause confusion in selecting necessary inputs and allocating results to other activities in the process. As an example, \emph{System} in the STPA terminology \cite{STPAhandbook} is equivalent to the vehicle function, or an \emph{item}, in the ISO 26262 terminology. This is one of the reasons why there are differences in literature regarding the mapping of STPA to activities in the automotive process.
For instance, NHTSA \cite{NHTSASOTIF} and SAE J3187 \cite{SAESTPA} mapped the identification of unsafe control actions (UCAs) to SOTIF hazard identification, however, Celik et al. took a different approach and mapped it to the safety concept \cite{asimsafecomp}.

\underline{\emph{Discussion:}} The STPA terminology shall be adapted to the automotive domain and then mapped to the relevant steps. For the mentioned example (mapping of UCA), the proposal of NHTSA \cite{NHTSASOTIF} and J3187 \cite{SAESTPA} only holds true if control actions are defined for an item as hazards exist at the item level. However, UCAs in both references are defined at the component or subsystem level. Our experience shows that UCA identification is related to activities in functional or technical safety concepts, depending on the abstraction level as also suggested by Celik et al. \cite{asimsafecomp}.

\vspace{0.3cm}
\begin{tabular}{|p{0.9\columnwidth}|}
  \hline
  \textbf{Challenge 2- Missing modular architecture (supported by literature and interviews):} \\  
  \hline
\end{tabular}
\vspace{0.1cm}

A modular architecture offers development costs reduction and maintainability improvement \cite{ModularKelly}.
A single-layered detailed design in a distributed development of a complex system increases the risk of development failure due to the involvement of teams and experts from diverse backgrounds \cite{SAESTPA}. 
Additionally, it may infringe on the intellectual property rights of both the OEMs and suppliers.

\underline{\emph{Discussion:}}
Although Abdulkhaleq et al.\cite{asimconti} discussed abstraction layers for STPA, it openly presents all details and modules as white boxes, looking at the most granular level. This setup requires shared IP and involvement of diverse expertise from all parties. 
This does not only lead to difficulties in performing the analysis but also makes the maintenance of the safety analysis challenging. In each iteration, all involved parties need to perform an impact analysis to ensure that changes in the new components do not affect their subsystem. 
For instance, in a study focused on safety requirements for a machine learning-based perception component \cite{asimsafecomp}, irrelevant subsystems such as the human-machine interface and actuators are also involved, which can distract the focus of analysis even if all relevant experts and teams are already involved. 
Additionally, full technical transparency is required in such a setup, which may not always be possible, especially in commercial projects and, as also mentioned by a participant in our interview study, only be possible for ``open source projects''.

In contrast, our interview study revealed that 12 out of 14  participants confirmed that there is no full technical transparency in the automotive industry between OEM and supplier. It also shows that 11 out of 14 participants confirmed that only the subsystem for which the supplier is responsible for shall be included in the analysis.
 As our findings indicate, there is a need for a black box modular approach in the STPA process for distributed development setups in the automotive industry. This is essential for setups with parties that are experts in specific technology and for protecting the intellectual property of each party.

\subsubsection{4.1.2 STPA Method for an OEM in Automotive System Engineering}

\label{subsec:solution OEM}
In the following, we are presenting the original STPA tailored to the automotive system engineering to overcome challenges originating from inconsistencies of terminologies mentioned in Sec. \ref{subsubsec:challenge OEM}.1.
The STPA consists of four steps\cite{STPAhandbook}, which are adapted to overcome challenge 1. For challenge 2, a modular approach will be proposed for step 2, which is partially used as well in ISO 21448\cite{sotif}.

\vspace{0.3cm}
\begin{tabular}{|p{0.9\columnwidth}|}
  \hline
  \textbf{Step 1 - Define the purpose of analyses:} \\  
  \hline
\end{tabular}
\vspace{0.1cm}

Step 1 involves defining the \underline{boundaries} of the item, function, or system and identifying the relevant \underline{losses} that shall be avoided. This study focuses exclusively on ``life or injury loss'' also known as functional safety and SOTIF addressed by the industry with applying ISO 26262 and ISO 21448. However, the scope can be expanded to include other losses.
The STPA handbook refers to hazards and constraints as ``system-level'', but in the automotive industry, where a vehicle function consists of multiple systems that each may contribute to multiple functions, it is more fitting to refer to them as ``(Vehicle-level) Hazards'' and ``Vehicle-level Safety Constraints'', as recommended by ISO 21448.

\underline{Hazards:} By combining the definitions of ISO 26262 \cite{ISO26262} and ISO 21448 \cite{sotif}, a hazard is caused by malfunctioning behavior (FuSa causes) or insufficiencies (SOTIF causes) of the item, which in combination with specific worst-case environmental conditions or operational situations (i.e., scenarios) would potentially lead to a loss. 

\underline{Vehicle-level constraints:} 
Looking at the formulation of the constraints in the example mentioned in SOTIF, they are in the same abstraction level and have the same formulation as the safety goal in the context of ISO 26262 \cite{SAESTPA} or validation target in ISO 21448. 

\begin{figure*}
\centering
  \includegraphics[width= 1\textwidth]{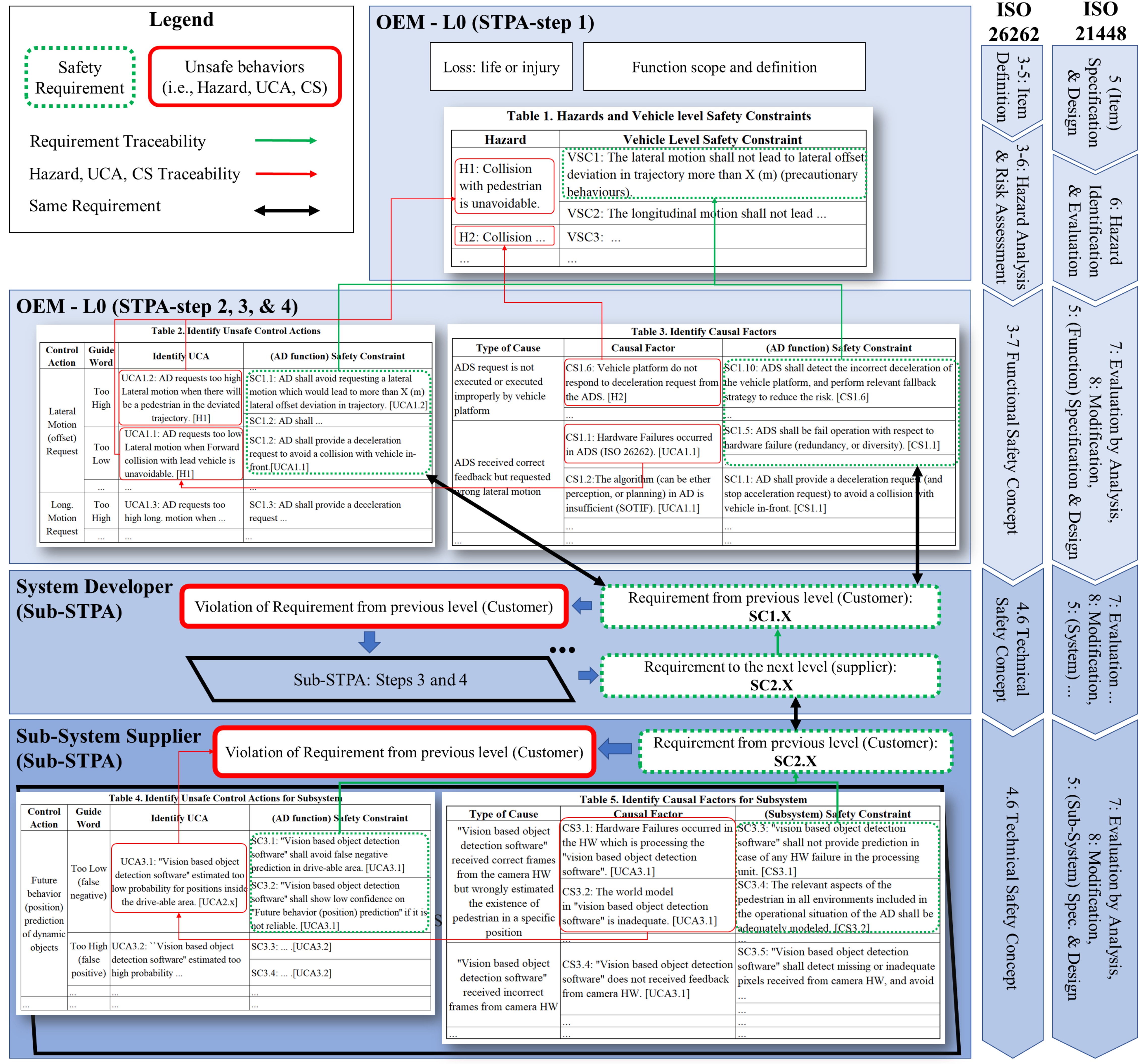}
  \caption{Represents the information flow, traceability of requirements, unsafe control actions and causal factors between the results of our proposed approach in each abstraction level. It also shows the mapping between each step and relevant activities in ISO 21448 and ISO 26262. A hypothetical example is provided to better show the process of applying STPA and Sub-STPA.}
  \label{fig:Traceability}
\end{figure*}

\vspace{0.3cm}
\begin{tabular}{|p{0.9\columnwidth}|}
  \hline
  \textbf{Step 2 - Model Control Structure:} \\  
  \hline
\end{tabular}
\vspace{0.1cm}

In Level 0, the AD system is treated as a black box, with a focus on the interaction between the system and stakeholders. This is a primary concern for OEMs and reducing complexity can aid in achieving this goal. Recent accidents involving the use or testing of AD\cite{uberaccident} have demonstrated that these incidents often result from interactions between the AD system and other stakeholders such as rule makers, OEMs, suppliers, and drivers.
In Fig.~\ref{fig:abstraction}, part B, the control structure of ``Unsupervised AD'' in L0 is depicted and the control actions are  ``Longitudinal Motion Request'' and ``Lateral Motion (offset) Request''

\vspace{0.3cm}
\begin{tabular}{|p{0.9\columnwidth}|}
  \hline
  \textbf{Step 3 - Identify Unsafe Control Actions:} \\  
  \hline
\end{tabular}
\vspace{0.1cm}

An Unsafe Control Action (UCA) is a state of a manner of control action, which in a specific context combined with worst-case environmental conditions leads to a hazard \cite{STPAhandbook}. Table 2 in Fig. \ref{fig:Traceability} illustrates the process of identifying UCAs for the example under consideration.
Each UCA leads to a violation of vehicle-level safety constraint. As it is shown in Fig. \ref{fig:Traceability}, Table 2, the UCAs shall be traceable to at least one hazard, as otherwise, the hazard list is incomplete and the relevant hazard shall be introduced.
The main purpose of any analysis is to identify the weaknesses of the system and then raise the need for a solution, which are formulated as \underline{Safety Requirements or constraints}. The relevant safety requirements to prevent UCAs are specified and presented in Table 2 in Fig. \ref{fig:Traceability}.

\vspace{0.3cm}
\begin{tabular}{|p{0.9\columnwidth}|}
  \hline
  \textbf{Step 4 - Identify Loss Scenario:} \\  
  \hline
\end{tabular}
\vspace{0.1cm}

After identifying the hazards in step 1 and UCAs in step 3, the corresponding causal factors shall be identified to be avoided by specifying safety constraints. The causal factors, which are related to the ``occurance of Unsafe Control Actions'', are traceable to a UCA, and the ones, which are the answer to ``improper or no execution of control actions'', are traceable directly to at least one Hazard. 
Both steps 3 and 4 should be in the same abstraction level as they analyze the same control structure and generate requirements for its components.
Like step 3, the relevant \underline{Safety Requirements or constraints} shall be specified to avoid causal factors.

\subsection{STPA Limitations for a Subsystem and Proposed adaptations:}

\label{subsec:challenge supplier}
As Koelln et al. \cite{COMPAD2} concluded, the usefulness of STPA is at vehicle level (L0) and if it is used for the next abstraction levels with more details, its advantages are not fully exploited.
Moreover, despite SAE J3187 acknowledging the abstraction levels (i.e., vehicle, system, subsystem), it proposes that more in-depth levels (i.e., subsystems) are better served by established analysis methodologies such as  Fault Tree Analysis (FTA) or Design Failure Mode Effect Analysis (DFMEA), rather than by STPA \cite{SAESTPA}.
This would inhibit suppliers of subsystems to use STPA and even if used, the traceability between the STPA results of the OEM and the supplier will not be systematically captured.

\subsubsection{4.2.1 Identified Challenges for subsystem supplier}
Challenges mentioned in Sec. \ref{subsubsec:challenge OEM}.1 are also applicable to suppliers. In the following, the supplier-specific challenges are presented:

\vspace{0.3cm}
\begin{tabular}{|p{0.9\columnwidth}|}
  \hline
  \textbf{Challenge 3- Hazard identification is not in the scope of sub-system supplier (supported by interview study):} \\ 
  \hline
\end{tabular}
\vspace{0.1cm}

Item/Function boundaries and Scenario are needed for performing the hazard analysis. As confirmed by 12 out of 14 participants, the subsystem supplier is not provided with the item or function definition, which is essential for defining the boundaries of item.
Moreover, a common issue reported by 8 out of 14 interviewees is that suppliers usually do not have access to the full list of scenarios, which results in an incomplete list of hazards.

\underline{\emph{Discussion:}} Missing item/Function boundaries, and Scenario leads to not being able to identify the hazards and vehicle-level constraints, which would lead to not being able to use STPA.
Inability to identify hazards and vehicle-level constraints is a consequence of missing item/function boundaries and scenarios as confirmed by the Interview study. This shortcoming inhibits subsystem suppliers to perform STPA.

\vspace{0.3cm}
\begin{tabular}{|p{0.9\columnwidth}|}
  \hline
  \textbf{Challenge 4- subsystem supplier cannot model the whole item (supported by literature and interview study):} \\ 
  \hline
\end{tabular}
\vspace{0.1cm}

To model the control structure, all subsystems and components including the scope of supplier shall be included. However, 11 out of 14 participants confirmed that a sub-system supplier only has access to the requirements and information relevant to the scope of supply and their subsystem design.
Based on our best knowledge and a thorough review of the literature, we have not come across any examples of analyses that were performed solely on a subsystem without modeling the rest of the system. Even in cases where the focus of the study is on a particular subsystem, such as the ``Machine Learning-Based Perception'' subsystem (or component) in \cite{asimsafecomp}, all involved systems in the control structure must be included, as required by the instruction of the control structure.

\underline{\emph{Discussion:}} Including all other subsystems involved in a function in the control structure may not always be efficient or even feasible due to the limited knowledge of a subsystem supplier about the whole system. 
In certain cases, including other subsystems can result in incorrect analyses as it may be based on incorrect assumptions or inadequate understanding of other subsystems.
For instance, by applying the STPA method, the control structure of a vision-based subsystem can be illustrated as shown in Fig.~\ref{fig:abstraction}, part C.
Comparing the control structure of the vision-based subsystem (8 elements) to that of the scope of supply (2 elements) reveals that analyzing the entire system requires approximately four times the effort and cost. Furthermore, the arbitration process in the sensor fusion subsystem involves numerous factors such as environmental conditions and object materials, which requires expertise beyond the scope of the vision-based subsystem. In addition, assuming these factors incorrectly could lead to a wrong result.

\vspace{0.3cm}
\begin{tabular}{|p{0.9\columnwidth}|}
  \hline
  \textbf{Challenge 5- No traceability between OEM's and supplier's STPA (supported by literature):} \\ 
  \hline
\end{tabular}
\vspace{0.1cm}

In the STPA handbook \cite{STPAhandbook}, the abstraction approach is proposed to handle the complexity at the start of the analysis. But then, each abstracted block will gradually be replaced by the expanded blocks, which leads to one single level of architecture without any black box module. A common theme among the studied literature is the use of a flat architecture, which does not require traceability between abstractions. 

\underline{\emph{Discussion:}}
To satisfy the need for traceability between safety analysis methods and requirements, ISO 26262 \cite{ISO26262} proposes a structure for safety requirements in four levels of abstraction in its ``Management of safety requirements'' section (Part 8, clause 6). It also illustrates the traceability between these levels, which has been expanded to cover ISO 21448 by Nouri et al. \cite{SEAACDDM}. 
A hierarchy of requirements, starting from safety goals to functional safety requirements, technical safety requirements and hardware and software safety requirements, requires at least three levels of safety analysis to create traceability to the parent requirement. For complex systems like AD, the abstraction levels can exceed five. However, this approach is not present in STPA, which can handle a maximum of two levels (L0 and L1) of abstraction.
However, while J3187 \cite{SAESTPA} and Abdulkhaleq et al. \cite{asimconti} both address abstraction levels, these levels are intended to gradually increase complexity within the same team or party, with the STPA analysis being refined and detailed design being included at each step as prescribed by the STPA handbook \cite{STPAhandbook}.

\subsubsection{4.2.2 Subsystem STPA (Sub-STPA) proposal} 

\label{subsec:solution supplier}
As outlined, there are three challenges that a subsystem supplier may face when using STPA, which could result in abandoning its use and failing to leverage its potential benefits.
The proposed adaptation includes specific instructions to overcome the aforementioned challenges and enable subsystem suppliers to perform STPA analysis with a focus only on their scope of supply.
Sub-STPA Step 1 aims to address challenge 3, while Sub-STPA Step 2 tackles both challenge 2 and 4. Fig. \ref{fig:abstraction} illustrates how Sub-STPA addresses challenge 5 and provides an example of the traceability between the results of STPA and Sub-STPA.

\vspace{0.3cm}
\begin{tabular}{|p{0.9\columnwidth}|}
  \hline
  \textbf{Sub-STPA Step 1:} \\ 
  \hline
\end{tabular}
\vspace{0.1cm}

The aforementioned instruction of ``Step 1'' is only applicable for the starting abstraction level that is named ``L0''. The goal of step 1 (i.e., ``purpose of the analysis'') is to define the safety constraints, which for levels other than L0 are already defined in the previous level as requirements in steps 3 or 4. 
The traceability and interactions between requirements and activities in each abstraction level as well as between levels are illustrated in Fig.~\ref{fig:Traceability}. In levels other than L0, Step 1 is replaced by a link to the safety requirements from the preceding abstraction levels, facilitating traceability from vehicle-level losses and hazards to the most detailed safety requirements in other abstraction levels.

\vspace{0.3cm}
\begin{tabular}{|p{0.9\columnwidth}|}
  \hline
  \textbf{Sub-STPA Step 2:} \\ 
  \hline
\end{tabular}
\vspace{0.1cm}

In contrast to step 2 of STPA, our approach focuses solely on the elements within the subsystem under analysis without requiring the inclusion of other subsystems or components.
In this adaptation, controlled process unit is responsible for providing inputs to the subsystem and receiving outputs from it. For example, in a vision subsystem, the output of the controlled process is the projection of light from the environment, while the input is the object information and predictions. Fig.~\ref{fig:abstraction} shows the difference between the modelings of control structure in STPA (part C) and Sub-STPA (part B) approaches. STPA requires inclusion of assumed abstraction of other subsystems, but in this proposal (part B), only the scope of supply is included in the analysis and the OEM (L0) is responsible for analyzing the entire subsystem.

\vspace{0.3cm}
\begin{tabular}{|p{0.9\columnwidth}|}
  \hline
  \textbf{Sub-STPA Step 3 and 4:} \\ 
  \hline
\end{tabular}
\vspace{0.1cm}

These 2 steps are the same as Sec. \ref{subsubsec:challenge OEM}.2. The results of the analysis for the vision subsystem are presented in Table 5 and 6 in Fig. \ref{fig:Traceability}, with traceability to the upper abstraction levels (L1 and L0).

\underline{\emph{Discussion:}}
During the interview, we initially provided the participants with a training session on STPA steps, using an example for an OEM. Next, we presented both STPA and Sub-STPA without indicating, which was the original STPA and which was our proposal, using an example for a vision-based object detection subsystem supplier (Tier 2). After presenting both approaches, we asked the participants ``which approach they would recommend for a Tier 2 to perform STPA, based on their experience''. Out of the 14 participants, 11 recommended Sub-STPA as an adaptation suitable for a subsystem supplier.

\section{Threats to Validity}

\label{sec:Threats}
\emph{Construction Validity:} 
As outlined in Sec.~\ref{sec:methodology}, the interview study is constructed to identify the challenges faced by performing STPA for both OEM and suppliers during automotive system engineering and to seek the recommended approach by the experts. 
The interview protocol underwent a peer review by the team and was further evaluated through a pilot study to assess its effectiveness.
In addition, we supported our reflections with evidence from industrial guidelines and current publications in the field. 

\emph{Internal Validity:} 
Although the experts are transparent in their opinion, we took two measures to prevent expert opinion from being influenced by interviewee opinions. First, we recorded the presentations of the method and the questions so that intonation or missing information would not affect the experts. Second, we anonymized the methods presented in question 4, omitting any specific names, and only presented the methods themselves along with an example to the interviewees. These measures helped to maintain the reliability of the results by ensuring that the experts' opinions were not influenced by any external factors.

Although all the experts in this study were selected from the Swedish industry, their extensive years of experience provided a diverse range of backgrounds regarding the companies and locations they had previously worked for. Moreover, as distributed development is not restricted to a single company, the participants' experiences extend beyond Sweden and they have worked with OEMs and suppliers from other regions of the world. Additionally, the safety standards that were the focus of this study are developed internationally.

\emph{External Validity:} 
Although this study is focused on safety in the automotive industry, particularly with regards to multi-abstraction levels, it is not limited to this field. The safety aspect of this method can also be applied to quality or cyber-security concerns.
Although we used an example in the context of autonomous driving, our proposed adaptation is not limited to any particular function or application, and can be applied in other industries such as avionics and defense.
The distributed development aspect is also not limited to the automotive industry and is more crucial in areas such as defense, where confidentiality is even more critical. The modular approach and multi-abstraction architecture can be extended to open-source projects as well, allowing teams to limit the scope of their work to their area of expertise and streamline the process.

\section{Conclusion \& Future Work}

\label{sec:conlusion}
Although some literature and practitioners question the applicability of STPA for subsystems, using an adaptation of STPA at the subsystem level is beneficial to maintain traceability to the most granular level. It serves as an enabler for approaches such as DevOps, facilitating the maintainability of the entire system, whether at the most abstract or granular level in the event of any changes.
This study identified five challenges faced by OEMs and suppliers in automotive context while performing STPA for a complex multi-abstraction architecture within a distributed development environment. Based on the findings, we modified the original STPA to fit OEM perspective and introduced  sub-STPA for the subsystem developers.
The results are validated by having confirmed the challenges by a literature study and interview study, where 11 out of 14 experts have recommended sub-STPA to be used by the subsystem developers.
Additionally, Fig. \ref{fig:Traceability} illustrates the achievement of traceability of STPA results across multiple levels of abstraction without any constraints on number of levels.

For future work, a detailed case study can be conducted to demonstrate the application of our proposed adaptation in a multilevel distributed system and integrating it with safety analysis methods. Although the present study does not aim to compare or integrate STPA with other inductive or deductive methods such as FMEA or FTA, we recognize it as a potential area for future research as a continuation of this study.

\section*{Acknowledgments}
Thanks to Vinnova (Diarienummer: 2021-02585), and WASP, for supporting this work. We are also grateful for valuable comments from Dr.~Beatriz Cabrero-Daniel.

\section*{Disclaimer}
The views and opinions expressed are those of the authors and do not necessarily reflect the official policy or position of Volvo Cars.

\end{document}